\documentclass[12pt,twoside]{article}

\usepackage{pslatex}    
\usepackage{amsfonts}
\usepackage{amsmath}
\usepackage{amsthm}
\usepackage{amscd}
\usepackage{cite}
\usepackage{epsf}
\usepackage{epsfig}
\usepackage{a4}

\def\beq{\begin{equation}}
\def\eeq{\end{equation}}
\def\bea{\begin{eqnarray}}
\def\eea{\end{eqnarray}}
\let\nn=\nonumber
\def\beann{\begin{eqnarray*}}
\def\eeann{\end{eqnarray*}}

\let\a=\alpha \let\be=\beta \let\g=\gamma 
\let\e=\varepsilon   

  \let\la=\lambda \let\m=\mu
 \let\x=\xi \let\p=\pi \let\r=\rho \let\s=\sigma
\let\om=\omega \let\ps=\psi
\let\ph=\varphi   
\let\Om=\Omega \let\Si=\Sigma 
 \let\G=\Gamma \let\D=\Delta

\newcommand{\bare}{\bar \varepsilon}

\let\qd=\quad \let\qqd=\qquad 

\def\epp{\, .}
\def\epc{\, ,}

\def\tst#1{{\textstyle #1}}

\theoremstyle{plain}

\newtheorem*{corollary*}{Corollary}
\newtheorem{conjecture}{Conjecture}

\theoremstyle{definition}

\def\2{\frac{1}{2}} \def\4{\frac{1}{4}}

\def\6{\partial}

\def\+{\dagger}

\def\<{\langle} \def\>{\rangle}

\def\i{{\rm i}}

\def\rd{{\rm d}}
\def\re{{\rm e}}

\DeclareMathOperator{\tr}{tr}

\DeclareMathOperator{\sign}{sign}
\DeclareMathOperator{\End}{End}

\def\Re{{\rm Re\,}} \def\Im{{\rm Im\,}}

\def\vv{\mathbf{v}}
\def\vs{\mathbf{s}}

\def\fa{\mathfrak{a}}

\def\faq{\overline{\mathfrak{a}}}

\def\aqq{\widetilde{\alpha}}
\def\bqq{\widetilde{\beta}}

\renewcommand{\appendix}{%
   \renewcommand{\section}{
        \secdef\Appendix\sAppendix}%
   \setcounter{section}{0}%
   \renewcommand{\thesection}{\Alph{section}}%
   \renewcommand{\theequation}{\thesection.\arabic{equation}}%
}

\newcommand{\Appendix}[2][?]{%
     \refstepcounter{section}%
     \setcounter{equation}{0}%
     \addcontentsline{toc}{appendix}%
          {\protect\numberline{\appendixname~\thesection} #1}%
     \vspace{\baselineskip}%
     {\noindent\large\bfseries\appendixname: #2\par}%
     \sectionmark{#1}\vspace{\baselineskip}}

\newcommand{\sAppendix}[1]{%
     {\noindent\large\bfseries\appendixname\:: #1\par}%
     \sectionmark{#1}\vspace{\baselineskip}}

\renewcommand{\tilde}{\widetilde}

\begin{document}

\thispagestyle{empty}

\begin{center}

{\Large {\bf Factorization of multiple integrals\\ representing the
density matrix\\ of a finite segment of the Heisenberg spin chain\\}}

\vspace{7mm}

{\large Herman E. Boos\footnote{
e-mail: boos@physik.uni-wuppertal.de. On leave from the Skobeltsyn
Institute of Nuclear Physics, Moscow.},
Frank G\"{o}hmann\footnote{e-mail: goehmann@physik.uni-wuppertal.de},
Andreas Kl\"umper\footnote{e-mail: kluemper@physik.uni-wuppertal.de}\\
and Junji Suzuki\footnote{e-mail: jszz@physik.uni-wuppertal.de.
On leave from Shizuoka University, Japan.}\\

\vspace{5mm}

Fachbereich C -- Physik, Bergische Universit\"at Wuppertal,\\
42097 Wuppertal, Germany\\}

\vspace{20mm}

{\large {\bf Abstract}}

\end{center}

\begin{list}{}{\addtolength{\rightmargin}{9mm}
               \addtolength{\topsep}{-5mm}}
\item
We consider the inhomogeneous generalization of the density matrix of a
finite segment of length $m$ of the antiferromagnetic Heisenberg chain.
It is a function of the temperature $T$ and the external magnetic
field $h$, and further depends on $m$ `spectral parameters' $\xi_j$.
For short segments of length 2 and 3 we decompose the known multiple
integrals for the elements of the density matrix into finite sums
over products of single integrals. This provides new numerically
efficient expressions for the two-point functions of the infinite
Heisenberg chain at short distances. It further leads us to conjecture
an exponential formula for the density matrix involving only a double
Cauchy-type integral in the exponent. We expect this formula to hold
for arbitrary $m$ and $T$ but zero magnetic field.
\\[2ex]
\end{list}

\clearpage

\section{Introduction}
The study of correlation functions of solvable
quantum systems took a new direction in 1992 when Jimbo et al.\
\cite{JMMN92} shifted the focus of attention from the two-point
functions to the density matrix of a sub-system as the object of
principal interest. They managed to derive an $m$-fold integral
expression for the density matrix of a chain segment of length $m$
of the infinitely long, antiferromagnetic $XXZ$-spin-$\2$ chain in the
off-critical regime. In the following years this work was generalized
to the critical regime \cite{JiMi96}, to non-zero magnetic field
\cite{KMT99b} and, most recently \cite{GKS05,GHS05}, to non-zero
magnetic field and temperature.

The above mentioned works have in common that they all deal with
an inhomogeneous generalization of the density matrix obtained
by placing `spectral parameters' or `inhomogeneities' onto $m$
consecutive vertical lines of the corresponding vertex model.
The homogeneous limit, when all the spectral parameters go to
zero, is involved and is usually only taken at a late stage of the
calculations after all formulae have been transformed into an
appropriate form. Moreover, the inhomogeneous density matrix
(at zero magnetic field and zero temperature) satisfies a first order
difference equation in the inhomogeneities which is part of the novel
reduced quantum Knizhnik-Zamolodchikov equation (rqKZ)
\cite{BJMST04a,BJMST04b}.

In \cite{BoKo01,BoKo02} the multiple integrals for a specific zero
temperature density matrix element of the isotropic Heisenberg chain
were explicitly evaluated for $m = 3$ and $m = 4$. This line of
research was further pursued in \cite{BKNS02,SSNT03}. In a second line
of research representations of the zero temperature density matrix
which do not involve multiple integrals were developed starting from
functional equations \cite{BKS03,BKS04a,BJMST04a,BJMST04b,BST05,SaSh05,%
SST05}. This helped to work out further concrete examples of short-%
range correlation functions, such that nowadays closed analytic
expressions for the ground state spin-spin correlators are known up to
the seventh neighbour. In its most elaborate form of the rqKZ equation
the functional equations were then used in \cite{BJMST05b} to derive
an exponential formula for the density matrix of a segment of length
$m$ of the XXX chain that involves only a double Cauchy-type integral
in the exponent. In the recent paper \cite{BJMST06} such a formula was
obtained for the XXZ and XYZ models as well.

This article is concerned with an attempt to generalize both of the
above approaches to finite temperature and partially also to finite
magnetic field. We concentrate on the isotropic Heisenberg chain
which on a periodic lattice of $L$ sites has the Hamiltonian
\begin{equation} \label{ham}
     H = J \sum_{j=1}^L \bigl( \s_j^\a \s_{j+1}^\a - 1 \bigr) \epp
\end{equation}
Here $\s_j^\a$, $\a = x, y, z$, acts as a Pauli matrix on site $j$
of the spin chain, and implicit summation over Greek indices is
understood. For this model we have achieved a factorization of
the density matrix for $m = 2$ and $m = 3$ into single integrals
valid for arbitrary finite temperature and finite magnetic field. 
Based on this result and on results obtained
from the high temperature expansion of the multiple integrals
\cite{TsSh05} we present a conjecture for a finite
temperature generalization of the exponential formula in
\cite{BJMST05b}.

The paper is organized as follows. In section 2 we recall the
multiple integral formula for the density matrix elements. 
In section 3 we explain the reduction procedure of those
integrals in the cases $m=2$ and $m=3$. In section 4 we
discuss our conjecture on the exponential form for zero magnetic
field. In the appendix we show compact formulae for the density
matrix elements for $m=3$. 

\section{The multiple integral formula for the density matrix}
The density matrix is a means to describe a sub-system as a part of a
larger system in thermodynamic equilibrium in terms of the degrees of
freedom of the sub-system. In our case the sub-system will consist of
$m$ consecutive sites of the spin chain. We first define a `statistical
operator' by
\begin{equation} \label{statoph}
      \r_L = \exp \bigl( - (H - h S^z)/T \bigr) \epc
\end{equation}
where $S^z = \2 \sum_{j=1}^L \s_j^z$ is the conserved $z$-component
of the total spin, $T$ is the temperature and $h$ is the homogeneous
external magnetic field. In terms of this statistical operator the
density matrix of a finite sub-chain of length $m$ of the infinite
chain is expressed as
\begin{equation} \label{defdensmatgen}
     D (T| h) = \lim_{L \rightarrow \infty}
                \frac{\tr_{m+1 \dots L} \r_L}
                     {\tr_{1 \dots L} \r_L} \epp
\end{equation}
By construction, the thermal average of every operator $A$ acting
non-trivially only on sites 1 to $m$ can then be written as
\begin{equation}
     \<A\>_{T, h} = \tr_{1 \dots m} \, A_{1 \dots m} D (T| h) \epc
\end{equation}
where $A_{1 \dots m}$ is the restriction of $A$ to the first $m$
lattice sites.

The tensor products $e^{\a_1}_{\be_1} \otimes \dots
\otimes e^{\a_m}_{\be_m}$ composed of $2 \times 2$ matrices
$e^\a_\be$, $\a, \be = 1, 2$, with a single non-zero entry at the
intersection of row $\be$ and column $\a$ form a basis of
$\End \bigl( ({\mathbb C}^2)^{\otimes m} \bigr)$. The matrix elements
of the density matrix with respect to this basis can be represented as
\begin{equation} \label{defdensmat}
     D^{\a_1 \dots \a_m}_{\be_1 \dots \be_m} (T| h) =
        \lim_{\x_1, \dots, \x_m \rightarrow 0}
        D^{\a_1 \dots \a_m}_{\be_1 \dots \be_m} (\x_1, \dots, \x_m) \epc
\end{equation}
where the expression under the limit is the inhomogeneous density
matrix element mentioned in the introduction. Due to the conservation
of $S^z$ it is non-zero only if $\sum_{j=1}^m (\a_j - \be_j) = 0$.

For the non-zero inhomogeneous density matrix elements
(\ref{defdensmat}) of the XXZ chain a multiple integral formula was
obtained in \cite{GKS05,GHS05}. When specialized to the isotropic
limit (\ref{ham}) it reads
\begin{align} \label{densint}
     D^{\a_1 \dots \a_m}_{\be_1 \dots \be_m} (\x_1, \dots, \x_m)
        = & \biggl[ \prod_{j=1}^{|\a^+|}
             \int_{\cal C} \frac{\rd \om_j}{2 \p (1 + \fa (\om_j))}
             \prod_{k=1}^{\aqq_j^+ - 1} (\om_j - \x_k - \i)
             \prod_{k = \aqq_j^+ + 1}^m (\om_j - \x_k) \biggr] \notag \\
          & \biggl[ \prod_{j = |\a^+| + 1}^{m}
             \int_{\cal C} \frac{\rd \om_j}{2 \p (1 + \faq (\om_j))}
             \prod_{k=1}^{\bqq_j^- - 1} (\om_j - \x_k + \i)
             \prod_{k = \bqq_j^- + 1}^m (\om_j - \x_k) \biggr]
             \notag \\ &
        \frac{\det G(\om_j, \x_k)}
             {\prod_{1 \le j < k \le m}
                 (\x_k - \x_j) ( \om_j - \om_k - \i)} \epp
\end{align}
The formula involves certain positive integers $\aqq_j^+$
and $\bqq_j^-$ derived from the sequences of indices $(\a_n)_{n=1}^m$,
$(\be_n)_{n=1}^m$ specifying the matrix element. The indices take
values $1, 2$ corresponding to spin-up or spin-down. We shall denote
the position of the $j$-th up-spin in $(\a_n)_{n=1}^m$ by $\a_j^+$ and
that of the $k$-th down-spin in $(\be_n)_{n=1}^m$ by $\be_k^-$. Then,
by definition, $\aqq_j^+ = \a_{|\a^+| - j + 1}^+$, where $|\a^+|$ is
the number of up-spins in $(\a_n)_{n=1}^m$, and $j = 1, \dots, |\a^+|$.
Similarly $\bqq_j^- = \be_{j - |\a^+|}^-$, $j = |\a^+| + 1, \dots, m$.

\begin{figure}
    \centering
    \includegraphics{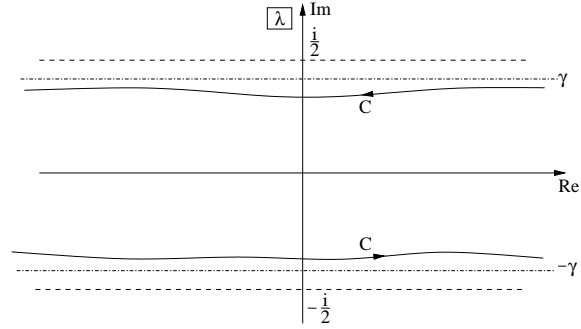}
    \caption{\label{fig:cancon} The canonical contour $C$ surrounds
             the real axis in counterclockwise manner inside the
             strip $- \2 < \Im \la < \2$.} 
\end{figure}

The functions $\fa (\om)$, $\faq (\om)$ and $G(\om, \x)$ are
transcendental and are defined as solutions of integral equations.
Through these functions and through the `canonical contour' $\cal C$
(see figure \ref{fig:cancon}) the temperature and the magnetic
field enter the multiple integral formula.
The integral equation for $\fa (\om)$ is non-linear,
\begin{equation} \label{nlie}
     \ln \fa(\la) = - \frac{h}{T} + \frac{2J}{\la (\la + \i) T}
                    - \int_{\cal C} \frac{\rd \om}{\p} \,
                      \frac{\ln (1 + \fa (\om))}
                           {1 + (\la - \om )^2} \epc
\end{equation}
and $\faq (\om) = 1/\fa (\om)$ by definition. Looking at the free
fermion limit $\D \rightarrow 0$ of the $XXZ$ chain \cite{GoSe05}
it is natural to interpret the combinations $1/(1 + \fa (\om))$
and $1/(1 + \faq (\om))$ as generalizations of the Fermi functions
for holes and particles, respectively, to the interacting case.

The function $G(\om, \x)$ is related to the magnetization density.
It satisfies the linear integral equation
\begin{equation} \label{ginteqn}
     G(\la,\x) + \frac{1}{(\la - \x) (\la - \x - \i)} =
                   \int_{\cal C} \frac{d \om}{\p (1 + \fa (\om))} \,
                   \frac{G (\om,\x)}{1 + (\la - \om)^2} \epp
\end{equation}
\section{Reduction of multiple integrals}
The multiple integrals (\ref{densint}) can be reduced by means of
the integral equation (\ref{ginteqn}). We demonstrate the method
with the example $m = 2$ and then merely present our results for
$m = 3$.
\subsection*{Emptiness formation probability for m = 2}
Consider (\ref{densint}) for $m = 2$ and $\a_1 = \a_2 = \be_1 = \be_2
= 1$. Then
\begin{multline}
     D^{11}_{11} (\x_1, \x_2) (\x_2 - \x_1) = \\
        \int_C \frac{d \om_1}{2\p (1 + \fa(\om_1))}
        \int_C \frac{d \om_2}{2\p (1 + \fa(\om_2))}
        \det \bigl( G(\om_j, \x_k) \bigr) \,
        \underbrace{\frac{(\om_1 - \x_1 - \i)(\om_2 - \x_2)}
                         {\om_1 - \om_2 - \i}}_{=: r(\om_1, \om_2)} \epp
\end{multline}
Here we may replace $r(\om_1, \om_2)$ by $[r(\om_1, \om_2) -
r(\om_2, \om_1)]/2$, since the domain of integration of the double
integral is symmetric. Now
\begin{multline}
     r(\om_1, \om_2) - r(\om_2, \om_1) = \\
        \frac{(\om_1 - \x_1 - \i)(\om_2 - \x_2)}{\om_1 - \om_2 - \i} +
        \frac{(\om_2 - \x_1 - \i)(\om_1 - \x_2)}{\om_1 - \om_2 + \i} =
        \frac{P(\om_1, \om_2)}{1 + (\om_1 - \om_2)^2} \epc
\end{multline}
and the polynomial $P(\om_1, \om_2)$ can be decomposed in such a way
that
\begin{equation}
     \frac{P(\om_1, \om_2)}{1 + (\om_1 - \om_2)^2} =
        \frac{p(\om_1) - p(\om_2)}{1 + (\om_1 - \om_2)^2}
        - \frac{2}{3} (\om_1 - \om_2)
\end{equation}
with
\begin{equation}
     p(\om) = \frac{2}{3} \om^3 - (\x_1 + \x_2 + \i) \om^2
              + \bigl[ \i (\x_1 + \x_2 + \tst{\frac{\i}{3}})
                                + 2 \x_1 \x_2 \bigr] \om \epp
\end{equation}
Then
\begin{multline}
     D^{11}_{11} (\x_1, \x_2) (\x_2 - \x_1) = \\
          \4 \sum_{P \in \mathfrak{S}^2} \sign(P)
          \int_C \frac{d \om_1 \, G(\om_1, \x_{P1})}
                      {\p (1 + \fa(\om_1))}
          \int_C \frac{d \om_2 \, G(\om_2, \x_{P2})}
                      {\p (1 + \fa(\om_2))}
          \biggl[ \frac{p(\om_1)}{1 + (\om_1 - \om_2)^2}
	        - \frac{2}{3} \om_1 \biggr] \epp
\end{multline}
The second term in the square brackets on the right is already of
factorized form. The first term can be reduced to factorized form
by means of the integral equation (\ref{ginteqn}). Finally
\begin{multline} \label{p22v1}
     D^{11}_{11} (\x_1, \x_2) (\x_1 - \x_2)
        = \sum_{P \in \mathfrak{S}^2} \sign(P)
          \biggl[ \frac{1}{12} (3 \x_{P1} - \x_{P2} + \i)
                               \phi_1 (\x_{P1})
                 + \frac{1}{6} \phi_2 (\x_{P2}) \\
                 - \frac{1}{6} \phi_1 (\x_{P1}) \phi_2 (\x_{P2})
                 - \frac{1}{24} (\x_{P1} - \x_{P2})
                   (1 + (\x_{P1} - \x_{P2})^2) \psi (\x_{P1}, \x_{P2})
                   \biggr] \epc
\end{multline}
where we have introduced a function
\begin{equation}
     \psi (\x_1, \x_2) =
        \int_C \frac{d \om}{\p (1 + \fa(\om))}
        \frac{G(\om, \x_1)}{(\om - \x_2)(\om - \x_2 - \i)}
\end{equation}
and a family of `moments'
\begin{equation}
        \phi_j (\x) = \int_C \frac{d \om \: \om^{j-1} G(\om, \x)}
                                 {\p (1 + \fa(\om))} \epc \qd
                                 j \in {\mathbb N} \epp
\end{equation}

Note that $\psi (\x_1, \x_2)$ is symmetric. This can be shown by means
of the integral equation (\ref{ginteqn}). The physical meaning of
$\psi (\x_1, \x_2)$ becomes evident in the limit of vanishing
temperature and magnetic field, where it can be expressed in terms
of gamma functions,
\begin{equation} \label{psizerozero}
     \lim_{T \rightarrow 0} \lim_{h \rightarrow 0} \ps (\x_1, \x_2)
        = 2 \i \: \6_x \ln \left[
            \frac{\G \bigl( \tst{\2 + \frac{\i x}{2}} \bigr)
                  \G \bigl( \tst{1 - \frac{\i x}{2}} \bigr)}
                 {\G \bigl( \tst{\2 - \frac{\i x}{2}} \bigr)
                  \G \bigl( \tst{1 + \frac{\i x}{2}} \bigr)}
                  \right]_{x = \x_1 - \x_2} \epp
\end{equation}
This is (up to a factor of $-2$) the two-spinon scattering phase
\cite{KuRe81,FaTa81}. It plays an important role in the recent works
\cite{BJMST04a,BJMST05b} as it is the only transcendental function
entering the general formula for the density matrix at zero
temperature and zero magnetic field. Instead of $\ps (\x_1, \x_2)$
we shall rather use the closely related expression
\begin{equation} \label{defgamma}
     \g (\x_1, \x_2) = \bigl[1 + (\x_1 - \x_2)^2\bigr]
                       \ps (\x_1, \x_2) - 1
\end{equation}
in terms of which our final formulae look neater. We also define
$\lim_{h \rightarrow 0} \g (\x_1, \x_2) =: \g_0 (\x_1, \x_2)$.

Considering the moments $\phi_j (\x)$ in the same limit of zero
temperature and magnetic field they turn into polynomials in $\x$
of order $j - 1$,
\begin{equation}
     \lim_{T \rightarrow 0} \lim_{h \rightarrow 0} \phi_j (\x)
        = \phi^{(0)}_j (\x)
        = (- \i \6_k)^{j-1} \, \frac{2 \re^{\i k \x}}{1 + \re^{k}}
          \Big|_{k=0} \epc
\end{equation}
for instance,
\begin{equation}
     \phi^{(0)}_1 (\x) = 1 \epc \qd
     \phi^{(0)}_2 (\x) = \x + \frac{\i}{2} \epc \qd
     \phi^{(0)}_3 (\x) = \x^2 + \i \x \epp
\end{equation}
These polynomials satisfy the difference equation
\begin{equation}
     \phi^{(0)}_j (\x) + \phi^{(0)}_j (\x - \i) = 2 \x^{j-1} \epp
\end{equation}
They allow us to define the `normalized moments',
\begin{equation}
     \ph_j (\x) = \phi_j (\x) - \phi^{(0)}_j (\x) \epc
\end{equation}
which vanish for $T, h \rightarrow 0$. We further introduce the
symmetric combinations
\begin{equation}
     \D_n (\x_1, \dots, \x_n) =
        \frac{\det (\ph_j (\x_k))\bigr|_{j, k = 1, \dots, n}}
             {\prod_{1 \le j < k \le n} \x_{kj}} \epc
\end{equation}
with the shorthand notation $\x_{kj} = \x_k - \x_j$.

The $\D_n$ will turn out to be particularly convenient for expressing
the density matrix elements for $m = 2, 3$. Using $\D_1$ and $\D_2$
in (\ref{p22v1}) we obtain
\begin{equation} \label{p22v2}
     D^{11}_{11} (\x_1, \x_2) = \frac{1}{4}
        + \frac{1}{4} (\D_1 (\x_1) + \D_1 (\x_2))
        + \frac{1}{6} \D_2 (\x_1, \x_2)
        - \frac{1}{12} \g (\x_1, \x_2) \epp
\end{equation}

The first moment $\ph_1$ is exceptional among the $\ph_j$ in that it
becomes trivial even for finite temperature if only the magnetic
field vanishes,
\begin{equation}
     \lim_{h \rightarrow 0} \ph_1 (\x) = 0 \epp
\end{equation}
It follows that
\begin{equation} \label{deltavan}
     \lim_{h \rightarrow 0} \D_j (\x) = 0 \epc \qd
        \text{for all $j \in {\mathbb N}$.}
\end{equation}
Thus, for vanishing magnetic field,
\begin{equation}
     D^{11}_{11} (\x_1, \x_2) = \frac{1}{4}
        - \frac{1}{12} \g_0 (\x_1, \x_2)
\end{equation}
and, in the homogeneous limit $\x_1, \x_2 \rightarrow 0$, we have
rederived the result
\begin{equation}
     \<\s_1^z\s_2^z\>_{T, h = 0} = 4 D^{11}_{11} (T|0) - 1
          = \frac{1}{3}
          - \lim_{h \rightarrow 0} \frac{1}{3}
	    \int_C \frac{d \om}{\p (1 + \fa(\om))}
            \frac{G(\om,0)}{\om (\om - {\rm i})}
\end{equation}
for the nearest-neighbour two-point function which alternatively can
be obtained \cite{GKS05b} by taking the derivative of the free energy
with respect to $1/T$. Here we can include the magnetic field into the 
calculation by simply taking the homogeneous limit in (\ref{p22v2}).
We obtain
\begin{equation}
     \<\s_1^z\s_2^z\>_{T, h} = 4 D^{11}_{11} (T|h) - 2 \D_1 (0) - 1
                             = \frac{2}{3} \D_2 (0, 0)
                             - \frac{1}{3} \g (0, 0)
\end{equation}
which seems to be a new result. Note that the magnetic field not only
enters through $\D_2$ but also through $\g$. The homogeneous limit
of $\D_n$ exists for all $n \in {\mathbb N}$ and is given by the
formula
\begin{equation}
     \lim_{\x_n \rightarrow 0} \dots \lim_{\x_1 \rightarrow 0}
          \D_n (\x_1, \dots, \x_n)
        = \det \Biggl[
          \frac{\6^{(k-1)}_\x \ph_j (\x)}{(k-1)!}
          \Biggr]_{\x = 0} \epp
\end{equation}
\subsection*{Complete density matrix for m = 2}
According to the rule $\a_1 + \a_2 = \be_1 + \be_2$ which reflects the
conservation of $S^z$ the density matrix for $m = 2$ has six
non-vanishing elements. Using the Yang-Baxter algebra and certain
identities of the type $D^1_1 (\x) + D^2_2 (\x) = 1$ we find four
independent relations between the six non-vanishing elements,
\begin{align} \label{m2densrel}
     & D^{12}_{12} (\x_1, \x_2)
        = D^1_1 (\x_1) - D^{11}_{11} (\x_1, \x_2) \epc \qd
       D^{21}_{21} (\x_1, \x_2)
        = D^1_1 (\x_2) - D^{11}_{11} (\x_1, \x_2) \epc \notag \\[1ex]
     & D^{22}_{22} (\x_1, \x_2) = D^{11}_{11} (\x_1, \x_2)
        - D^1_1 (\x_1) - D^1_1 (\x_2) + 1 \epc \notag \\[1ex]
     & D^{21}_{12} (\x_1, \x_2) - D^{12}_{21} (\x_1, \x_2) =
       \frac{D^1_1 (\x_1) - D^1_1 (\x_2)}{\i \x_{12}} \epp
\end{align}
Thus, we have to calculate one more independent linear combination,
say $D^{21}_{12} (\x_1, \x_2) + D^{12}_{21} (\x_1, \x_2)$, in order
to determine the complete density matrix for $m = 2$. The calculation
can be done along the lines described above.

In order to obtain a convenient description of all density matrix
elements we shall resort to a notation that we borrowed from
\cite{BJMST06}\footnote{This definition was first introduced in
\cite{BJMST04a} and later modified in \cite{BJMST06}.}. We arrange
them into a column vector $h_m \in ({\mathbb C}^2)^{\otimes 2 m}$ with
coordinates labeled by $+, -$ instead of $1, 2$ according to the rule,
\begin{equation} \label{vecmat}
     h_m^{\e_1, \dots, \e_m, \bare_m, \dots, \bare_1}
         (\la_1, \dots, \la_m)
        = D^{(3 - \e_1)/2,  \dots, (3 - \e_m)/2}_%
            {(3 + \bare_1)/2, \dots, (3 + \bare_m)/2}
            (\xi_1, \dots, \xi_m) \cdot
            \prod_{j = 1}^m (- \bare_j) \epc
\end{equation}
where $\la_j = - \i \x_j$ for $j = 1, \dots, m$.

Then, setting $\vv^t = (h^{++--}_2, h^{+-+-}_2, h^{+--+}_2, h^{-++-}_2,
h^{-+-+}_2, h^{--++}_2)$,
\begin{multline}
     \vv = \frac{1}{4} \vv_0
         - \frac{1}{12} \g(\x_1, \x_2) \vv_1 \\
         + \frac{1}{4} (\D_1 (\x_1) + \D_1 (\x_2)) \vv_2
         - \frac{1}{4} (\D_1 (\x_1) - \D_1 (\x_2)) \vv_3
         + \frac{1}{6} \D_2 (\x_1, \x_2) \vv_4
\end{multline}
with
\begin{align}
     \vv_0 & = (1, -1, 0, 0, -1, 1)^t \epc \notag \\[1ex]
     \vv_1 & = (1, 1, -2, -2, 1, 1)^t \epc \notag \\[1ex]
     \vv_2 & = (1, 0, 0, 0, 0, -1)^t \epc \notag \\[1ex]
     \vv_3 & = (0, 1, \i \x_{12}^{-1}, -\i \x_{12}^{-1}, -1, 0)^t
               \epc \notag \\[1ex]
     \vv_4 & = (1, 1, 1, 1, 1, 1)^t \epp
\end{align}
From this we can read off the transverse neighbour correlation functions
in a magnetic field to be $\<\s_1^x \s_2^x\>_{T,h} = - \frac{1}{3}
\D_2 (0,0) - \frac{1}{3} \g (0,0)$.

\subsection*{Emptiness formation probability for m = 3}
As in the case $m=2$ we can reduce the triple integral (\ref{densint})
representing the emptiness formation probability for $m = 3$ to
sums over products of single integrals. Again we have to decompose
the rational functions in the integrand appropriately and use the
integral equation (\ref{ginteqn}). It turns out that the final result
can be represented in terms of the functions $\g$ and $\D_j$, $j =
1, 2, 3$,
\begin{multline} \label{p3}
     D^{111}_{111} (\x_1, \x_2, \x_3) = \\[2ex] \frac{1}{24}
        + \frac{1 + 5 \x_{12} \x_{13}}{40 \x_{12} \x_{13}} \D_1 (\x_1)
        + \frac{1 + 2 \x_{13} \x_{23}}{24 \x_{13} \x_{23}}
          \D_2 (\x_1, \x_2) + \frac{1}{60} \D_3 (\x_1, \x_2, \x_3)
          \\[2ex]
        + \frac{1 - \x_{13} \x_{23}}{24 \x_{13} \x_{23}} \,
          \g (\x_1, \x_2)
        - \frac{3 + 2 \x_{12}^2 + 5 \x_{13} \x_{23}}
               {120 \x_{13} \x_{23}} \, \g (\x_1, \x_2) \D_1 (\x_3)
               \\[2ex]
        + \text{cyclic permutations.}
\end{multline}
In the limit of vanishing magnetic field (\ref{deltavan}) applies
and our result reduces to
\begin{equation} \label{p3h0}
     D^{111}_{111} (\x_1, \x_2, \x_3) = \frac{1}{24}
        + \frac{1 - \x_{13} \x_{23}}
               {24 \x_{13} \x_{23}} \g_0 (\x_1, \x_2)
                    + \text{cyclic permutations.}
\end{equation}
Note that the only effect of taking the limit $T \rightarrow 0$ is that
the function $\g_0 (\x_1, \x_2)$ changes into its zero temperature form 
(\ref{psizerozero}), (\ref{defgamma}).

\subsection*{Complete density matrix for m = 3}
We have factorized the full density matrix for $m = 3$. It contains 20
non-vanishing matrix elements, each of similar form to that in
(\ref{p3}). Except for the emptiness formation probability
$D^{111}_{111}$ we also reduced the integrals for the symmetric
combination $D^{112}_{121} + D^{121}_{211} + D^{211}_{121} +
D^{112}_{211} + D^{121}_{112} + D^{211}_{112}$. Using relations similar
to (\ref{m2densrel}) and the high temperature expansion data for the
inhomogeneous density matrix elements up to the order $T^{-3}$ this was
then enough to conjecture the complete density matrix for $m=3$. We
show it in a compact notation in the appendix. For our purposes here
it is sufficient to know that all density matrix elements are linear
combinations of the functions $\D_1$, $\D_2$, $\D_3$, $\g$ and
$\g \D_1$ with coefficients rational in the differences $\x_{jk}$. For
the case of vanishing magnetic field an `exponential formula' for all
density matrix elements is suggested in the next section.

\subsection*{The next-to-nearest neighbour two-point functions}
It is a rather straightforward exercise to work out the
next-to-nearest neighbour two-point functions from our general
result for the $m = 3$ density matrix. We have to take the
appropriate linear combinations of density matrix elements and
have to carry out the homogeneous limit $\x_1, \x_2, \x_3
\rightarrow 0$. We obtain, for instance,
\begin{subequations}
\label{szsx3}
\begin{align}
    \<\s_1^z \s_3^z\>_{T,h} & = \frac{2}{3} \D_2 (0,0)
                        - \frac{1}{3} \g (0,0)
                          \notag \\ & \mspace{36.mu}
                        - \frac{1}{6} (\D_2)_{xx} (0,0)
                        + \frac{1}{3} (\D_2)_{xy} (0,0) 
                        - \frac{1}{6} \g_{xx} (0,0) 
                        + \frac{1}{3} \g_{xy} (0,0) \epc \\[1ex]
    \<\s_1^x \s_3^x\>_{T,h} & = - \frac{1}{3} \D_2 (0,0)
                        - \frac{1}{3} \g (0,0) 
                          \notag \\ & \mspace{36.mu}
                        + \frac{1}{12} (\D_2)_{xx} (0,0) 
                        - \frac{1}{6} (\D_2)_{xy} (0,0) 
                        - \frac{1}{6} \g_{xx} (0,0) 
                        + \frac{1}{3} \g_{xy} (0,0) \epp
\end{align}
\end{subequations}
Here we denoted derivatives with respect to the first and second
argument, respectively,  by subscripts $x$ and $y$. Equations
(\ref{szsx3}) generalize an important result of Takahashi
\cite{Takahashi77} to include the temperature and the magnetic field.

\section{The exponential formula}
We observed in the previous section that for $m = 2, 3$ the density
matrix for zero magnetic field is determined by a single transcendental
function $\g_0$ (recall that $\lim_{h \rightarrow 0} \D_j = 0$). The
situation is the same as for zero temperature. In fact, even the
coefficients agree. Hence, it is tempting to substitute the function
$\g_0$ for its zero temperature analogue into the general exponential
formula recently obtained in \cite{BJMST05b}. This formula then gives
the correct result for the 6 non-trivial density matrix elements for
$m = 2$ and also for the emptiness formation probability and for the
symmetric combination of density matrix elements mentioned in the
previous section for $m = 3$. It further coincides up to order
$T^{-3}$ with the high-temperature expansion data for all 20
non-vanishing inhomogeneous density matrix elements for $m = 3$.
For $m = 4$ we compared the conjectured form in the homogeneous case
with the high-temperature expansion obtained from the homogeneous
version \cite{GKS05,GHS05} of the multiple integral formula
(\ref{densint}). For the emptiness formation probability we found full
agreement up to the order of $T^{-11}$.
\begin{conjecture}
The density matrix of a finite sub-chain of length $m$ of the infinite
XXX Heisenberg chain at finite $T$ (for $h = 0$) is determined by the
vector
\begin{align} \label{expform}
     h_m (\la_1, \dots, \la_m)
        & = \frac{1}{2^m} \re^{\Om^T_m (\la_1, \dots, \la_m)} \vs_m
            \epc \qqd \vs_m = \prod_{j=1}^m s_{j, \bar j} \epc
            \\[1ex] \label{defom}
     \Om^T_m (\la_1, \dots, \la_m) & = \frac{(-1)^{(m-1)}}{4}
        \int \int \frac{d \m_1}{2 \p \i} \frac{d \m_2}{2 \p \i}
        \frac{\g_0 (\i \m_1, \i \m_2) (\m_1 - \m_2)}
             {[1 - (\m_1 -\m_2)^2]^2} \\ & \notag \mspace{-72.mu} \times
        \tr_{\m_{1,2},2,2} \Bigl\{
           T \bigl( \tst{\frac{\m_1 + \m_2}{2}}; \la_1, \dots, \la_m
                      \bigr) \otimes \bigl[
           T (\m_1; \la_1, \dots, \la_m) \otimes
           T (\m_2; \la_1, \dots, \la_m) {\cal P}^- \bigr] \Bigr\} \epc
\end{align}
through (\ref{vecmat}). By the integral over $\m_1$, $\m_2$ it is
meant to take the residues at the poles $\la_1, \dots, \la_m$ of the
integrand.
\end{conjecture}
For the notation we are referring to \cite{BJMST05b}\footnote{In fact,
the only difference between our formula (\ref{expform}), (\ref{defom})
and the result of \cite{BJMST05b} is in the function $\g_0$. In
\cite{BJMST05b} a function $\om$ was used which is related to $\g_0$ by
\[
     \omega(\la_1-\la_2) = \lim_{T\rightarrow 0}
                           \frac{\g_0 (i\la_1,i\la_2)}
			        {2\;(1-(\la_1-\la_2)^2)} \epp
\]
}: The vector
$s = \binom{1}{0} \otimes \binom{0}{1} - \binom{0}{1} \otimes
\binom{1}{0}$ is the spin singlet in ${\mathbb C}^2 \otimes
{\mathbb C}^2$. The vector spaces in $({\mathbb C}^2)^{\otimes 2m}$
are numbered in the order $1, 2, \dots, n, \bar n, \overline{n - 1},
\dots, \bar 1$. This defines $\vs_m$. ${\cal P}^-$ is the projector
onto the one-dimensional subspace of ${\mathbb C}^2 \otimes
{\mathbb C}^2$ spanned by $s$.

In order to define the transfer matrices in the integrand in
(\ref{defom}) we first of all introduce an $L$-matrix $L (\la) \in
U (\mathfrak{sl}_2) \otimes \End {\mathbb C}^2$,
\begin{equation} \label{defl}
     L(\la) = \frac{\r (\la, d)}{2 \la + d}
             (2 \la + 1 + \Si^\a \otimes \s^\a) \epc
\end{equation}
where the $\Si^\a \in \mathfrak{sl}_2$ are a basis satisfying $[\Si^\a,
\Si^\be] = 2 \i \e^{\a \be \g} \Si^\g$, where $d$ is determined
by the Casimir element through $d^2 = (\Si^\a)^2 + 1$ and where
$\r (\la, d)$ satisfies the functional relation
\begin{equation}
     \r(\la, d) \r(\la - 1, d) = \frac{2 - 2 \la - d}{2 \la - d}
\end{equation}
(for more details see \cite{BJMST05b}). Then, for integer $z$, the
`transfer matrices'
\begin{multline}
     \tr_z T(\la; \la_1, \dots, \la_n) = \\
     \tr_z L_{\bar 1} (\la - \la_1 - 1) \dots
           L_{\bar n} (\la - \la_n - 1) L_n (\la - \la_n) \dots
           L_1 (\la - \la_1)
\end{multline}
entering (\ref{defom}) are defined by substituting the irreducible
representation of $U (\mathfrak{sl}_2)$ of dimension $z$ into the
definition (\ref{defl}) of the $L$-matrices. For non-integer
$z$ this can be analytically continued into the complex plane.

\section{Discussion}
Starting from the multiple-integral formula (\ref{densint}) we have
investigated the density matrix of a finite segment of length $m$
of the infinite isotropic Heisenberg chain at finite temperature and
finite magnetic field. We found that the multiple integrals can be
reduced to sums over products of single integrals, in much the same way
as for $T, h = 0$. On the one hand this gives new efficient formulae
for the calculation of finite-temperature short-range correlations 
of the XXX chain in the thermodynamic limit. On the other hand
this shows that the density matrix and the correlation functions
of the Heisenberg chain at finite temperature and finite magnetic
field may be explored in a similar manner and to much the same extent
as in the ground state case without magnetic field. When the magnetic
field is switched off, but the temperature is kept finite the
`algebraic structure' of the density matrix, as it shows up in the
rational functions in the differences of the spectral parameters,
seems to be still the same as for zero temperature, and only the
`physical part' of the expressions, encoded in the transcendental
function $\g_0$, changes. This is certainly true for $m \le 2$ and
for some of the density matrix elements for $m = 3$. From our
high-temperature analysis it seems most likely true also for all
density matrix elements for $m = 3, 4$, whence our conjecture 1 in the
previous section. A second conjecture we are tempted to formulate
inspecting our results for small $m$ is the following: {\it In the
case of non-vanishing magnetic field the density matrix elements for
a segment of length $m$ seem to depend only on $\g$ and on $\D_1,
\dots, \D_m$.}
\\[1ex]{\bf Acknowledgement.}
The authors are indebted to M. Shiroishi and Z. Tsuboi for providing
their unpublished high-temperature expansion data. They would like
to thank M. Jimbo, T. Miwa, M. Shiroishi, F. Smirnov, M. Takahashi,
Y. Takeyama and Z. Tsuboi for stimulating discussions. HB is supported
by the RFFI grant \# 04-01-00352. JS acknowledges financial support
by the DFG-funded research training group 1052 -- `representation
theory and its applications' and by the Ministry of Education of Japan
through a Grand-in-Aid for Scientific Research \# 14540376.

\def\({\left(}
\def\){\right)}
\newcommand{\bra}[1]{\langle #1 |}        
\newcommand{\ket}[1]{{| #1 \rangle}}      
\newcommand{\br}[1]{{\langle #1 \rangle}}  
\newcommand{\ds}[1]{\displaystyle #1}
\renewcommand{\Re}{\mathop{\rm Re}}  

\newcommand{\ena}{\end{eqnarray}}
\newcommand{\ba}{\begin{array}}
\newcommand{\ea}{\end{array}}
\renewcommand{\o}[1]{\overline{#1}\,}
\def\og{\overline{\gamma}}
\def\P{\mathcal P}

\newcommand{\R}{{\mathbb R}}
\newcommand{\C}{{\mathbb C}}
\newcommand{\Z}{{\mathbb Z}} 
\newcommand{\Q}{{\mathbb Q}} 
\newcommand{\A}{{\mathcal A}} 
\newcommand{\cP}{\mathcal{P}}
\newcommand{\cC}{\mathcal{C}}
\newcommand{\cL}{\mathcal{L}}
\newcommand{\cK}{\mathcal{K}}
\newcommand{\Xh}{\widehat{X}}
\newcommand{\Oh}{\widehat{\Omega}}
\newcommand{\Gh}{\hat{G}}
\newcommand{\Rc}{\check{R}}
\newcommand{\Oc}{\overset{\circ}{\Omega}}
\newcommand{\slt}{\mathfrak{sl}_2}
\newcommand{\wt}{{\rm wt}\,}
\newcommand{\Tr}{{\rm Tr}}
\newcommand{\vac}{{\rm vac}}
\newcommand{\cJ}{J^{(c)}}
\newcommand{\Hom}{\mathop{\rm Hom}}
\newcommand{\Ker}{\mathop{\rm Ker}}
\renewcommand{\Im}{\mathop{\rm Im}}
\newcommand{\isoto}[1][]%
{{\mathop{\buildrel{\sim}\over\longrightarrow}\limits_{#1}}}
\newcommand{\To}[1][\phantom{aaaa}]{\xrightarrow{\,#1\,}}

\newcommand{\al}{\alpha}
\newcommand{\eb}{\bar{\epsilon}}
\newcommand{\ve}{\varepsilon}
\newcommand{\bs}{\mathbf{s}}
\newcommand{\ot}{\tilde{\omega}}
\newcommand{\hti}{h^{Corr}}
\newcommand{\pt}{\tilde{p}}
\newcommand{\rt}{\tilde{\rho}}
\newcommand{\Ft}{\tilde{G}}
\newcommand{\F}{{G}}
\newcommand{\Kb}{{\bf K}}
\newcommand{\M}{{\bf M}}
\newcommand{\ad}{\mathop{\rm ad}}
\newcommand{\V}{\mathcal{V}}

{\appendix
\Appendix{Density matrix elements for m = 3}
\noindent
Here we show our result for other elements of the density matrix in the
$m=3$ case. We consider the 3 by 3 block  
\bea
D_{3\times 3}=
\begin{pmatrix}
D_{112}^{112} & D_{121}^{112} & D_{211}^{112}\\
D_{112}^{121} & D_{121}^{121} & D_{211}^{121}\\
D_{112}^{211} & D_{121}^{211} & D_{211}^{211}
\label{D33}
\end{pmatrix} \epp
\ena
Another 3 by 3 block can be obtained from (\ref{D33}) by the
substitution $1\leftrightarrow 2$ and $h\rightarrow -h$. The answer
looks as follows
\begin{align}
&
D_{3\times 3}=
A_0 + \sum_{1\le i<j\le 3}A_{i,j}\gamma_{i,j}
+\sum_{1\le i<j\le 3;k\ne i,j}
A_{i,j|k}\gamma_{i,j}\Delta_1(\xi_k)\nn\\
&+\sum_{i=1}^3 B_i\Delta_1(\xi_i)
+\sum_{1\le i<j\le 3}B_{i,j}\Delta_2(\xi_i,\xi_j)
+B_{1,2,3}\Delta_3(\xi_1,\xi_2,\xi_3) \epc
\label{Dres}
\end{align}
where $\gamma_{i,j}=\gamma(\xi_i,\xi_j)$ and $A_0, A_{i,j}, A_{i,j|k},
B_i, B_{i,j}, B_{1,2,3}$ are 3 by 3 matrices. In order to define them
let us introduce 3 by 3 matrices with the following elements
\begin{align}
& (e_0)_{k,l} = \delta_{k,l}, \quad (Z_0)_{k,l} = 1, \quad 
(X_i)_{k,l} = 
\begin{cases}
-1 & \text{if $k=l=i$}\\
\;\;\;1 & \text{if $k=l\ne i$}\\
\;\;\;0 & \text{otherwise}
\end{cases},
\nn\\
& X^{\pm}_{i,j} = e_i^j\pm e_j^i\quad\text{for $i<j$},
\quad (e_i^j)_{k,l} = \delta_{i,k}\delta_{j,l},
\quad\quad 1\le i,j,k,l \le 3 \epp
\label{X}
\end{align}
We also define $X^{\pm}_{j,i}=X^{\pm}_{i,j}$ and 
$X_{i+3n}=X_{i},\;\; X^{\pm}_{i+3n,j+3m}=X^{\pm}_{i,j}$
for $n,m\in \Z$. Then
\begin{align}
A_0=&\frac18 e_0, \notag\\
A_{i,j} =& -\frac1{24}\,\frac1{\xi_{ik}\xi_{jk}}\; (X_2-2 X^+_{1,3}) +
\frac{\i}{24}\;\biggl(\frac{(-1)^j}{\xi_{ik}}+
\frac{(-1)^i}{\xi_{jk}}\biggr)\;Y +\frac{1}{24}\;(X_{-j} -2 X^+_{-i,-k}),\notag\\
Y = & X^-_{1,2}-X^-_{1,3}+X^-_{2,3} \epc \notag
\label{A}\\
A_{i,j|k} =& \frac1{24}\;(e_0-2 X^+_{-i,-k})
-\frac{1}{24}\;\frac{1}{\xi_{ik}\xi_{jk}}\;
\biggl(\frac25 Z_0 - e_0\biggr) +
\frac{1}{24}\;
\biggl(\frac{\i}{\xi_{ik}}+\frac{\i}{\xi_{jk}}\biggr)\;X^-_{1,3}\\ 
&- \frac{1}{24}\;\frac{\xi_{ij}}{\xi_{ik}}\;
\biggl(\frac25 Z_0 -X^+_{-j,-i}-X^+_{-i,-k}\biggr)
+\frac{1}{24}\;\frac{\xi_{ij}}{\xi_{jk}}\;
\biggl(\frac25 Z_0 -X^+_{-j,-k}-X^+_{-i,-k}\biggr) \epp \notag
\end{align}
\begin{align}
B_i=&\frac18 X_{-i+1}+\frac18\;\frac{\i}{\xi_{ij}}\;X^-_{-i,-k}
+\frac18\;\frac{\i}{\xi_{ik}}\;X^-_{-j,-k}-
\frac{1}{8}\;\frac{1}{\xi_{ij}\xi_{ik}}\;\biggl(\frac15 Z_0-X^+_{1,3}\biggr) \epc \notag\\
B_{i,j}=&-\frac1{12}\;\bigl(X_{-j}+X^+_{-i,-k}\bigr)
-\frac{1}{24}\;\frac{1}{\xi_{ik}\xi_{jk}}\;\bigl(X_2 + X^+_{13}\bigr)\notag\\
&+\frac1{24}\;\frac{\i}{\xi_{ik}}\;
\biggl(2 X^-_{-j,-k}-(-1)^k X^-_{-i,-j}-(-1)^i X^-_{-i,-k}\biggr)
\notag\\
&+\frac1{24}\;\frac{\i}{\xi_{jk}}\;
\biggl(2 X^-_{-i,-j}-(-1)^k X^-_{-j,-k}-(-1)^j X^-_{-i,-k}\biggr)\epc \notag\\
B_{1,2,3}=&-\frac1{20}\;Z_0 \epp
\label{B}
\end{align}
It is implied in the above formulae that the triple $(i,j,k)$ 
is always the cyclic permutation of $(1,2,3)$ and also that
$A_{i,j}=A_{j,i}, A_{i,j|k}=A_{j,i|k}, B_{i,j}=B_{j,i}$.
Let us also mention that the matrices $A_0$ and $A_{i,j}$ coincide 
with those which can be obtained from the formula (\ref{defom}).
}


\begin{thebibliography}{10}

\bibitem{BJMST05b}
H.~Boos, M.~Jimbo, T.~Miwa, F.~Smirnov, and Y.~Takeyama, \emph{Density matrix
  of a finite sub-chain of the {H}eisenberg anti-ferromagnet}, 
  to appear in Lett. Math. Phys.,  hep-th/0506171, 2005.

\bibitem{BJMST04a}
\bysame, \emph{A recursion formula for the correlation functions of an
  inhomogeneous {XXX} model}, Algebra and Analysis \textbf{17} (2005), 115.

\bibitem{BJMST04b}
\bysame, \emph{Reduced $q${KZ} equation and correlation functions of the {XXZ}  model}, Comm. Math. Phys. \textbf{261} (2006), 245.

\bibitem{BJMST06}
\bysame, \emph{Algebraic representation of correlation functions
in integrable spin chains}, to appear in 
the Annales Henri Poincare, volume dedicated to D. Arnaudon,
hep-th/0601132, 2006.

\bibitem{BoKo01}
H.~E. Boos and V.~E. Korepin, \emph{Quantum spin chains and {R}iemann zeta
  function with odd arguments}, J. Phys. A \textbf{34} (2001), 5311.

\bibitem{BoKo02}
\bysame, \emph{Evaluation of integrals representing correlations in the {XXX}
  {H}eisenberg spin chain}, MathPhys Odyssey 2001 -- Integrable Models and
  Beyond -- In Honor of Barry M. McCoy (M.~Kashiwara and T.~Miwa, eds.),
  Birkh\"auser, Boston, 2002, Progress in Mathematical Physics, Vol.\ 23,
  pp.~65--108.

\bibitem{BKNS02}
H.~E. Boos, V.~E. Korepin, Y.~Nishiyama, and M.~Shiroishi, \emph{Quantum
  correlations and number theory}, J. Phys. A \textbf{35} (2002), 4443.

\bibitem{BKS03}
H.~E. Boos, V.~E. Korepin, and F.~A. Smirnov, \emph{Emptiness formation
  probability and quantum {K}nizhnik-{Z}amolodchikov equation}, Nucl. Phys. B
  \textbf{658} (2003), 417.

\bibitem{BKS04a}
\bysame, \emph{New formulae for solutions of quantum {K}nizhnik-{Z}amolodchikov
  equation on level $-4$}, J. Phys. A \textbf{37} (2004), 323.

\bibitem{BST05}
H.~E. Boos, M.~Shiroishi, and M.~Takahashi, \emph{First principle approach to
  correlation functions of spin-1/2 {H}eisenberg chain: fourth-neighbor
  correlators}, Nucl. Phys. B \textbf{712} (2005), 573.

\bibitem{FaTa81}
L.~D. Faddeev and L.~A. Takhtajan, \emph{Spectrum and scattering of excitations
  in the one-dimensional isotropic {Heisenberg} model}, Zap. Nauchn. Sem. LOMI
  \textbf{109} (1981), 134, translated in J. Soviet Math.\ {\bf 24} (1984) 241.

\bibitem{GHS05}
F.~G\"ohmann, N.~P. Hasenclever, and A.~Seel, \emph{The finite temperature
  density matrix and two-point correlations in the antiferromagnetic {XXZ}
  chain}, J. Stat. Mech. (2005), P10015.

\bibitem{GKS05b}
F.~G\"ohmann, A.~Kl\"umper, and A.~Seel, \emph{Emptiness formation probability
  at finite temperature for the isotropic {H}eisenberg chain}, Physica B
  \textbf{359-361} (2005), 807.

\bibitem{GKS05}
\bysame, \emph{Integral representation of the density matrix of the {XXZ} chain
  at finite temperature}, J. Phys. A \textbf{38} (2005), 1833.

\bibitem{GoSe05}
F.~G\"ohmann and A.~Seel, \emph{{XX} and {I}sing limits in integral formulae
  for finite temperature correlation functions of the {XXZ} chain}, Theor.
  Math. Phys. \textbf{146} (2006), 119.

\bibitem{JMMN92}
M.~Jimbo, K.~Miki, T.~Miwa, and A.~Nakayashiki, \emph{Correlation functions of
  the {XXZ} model for {$\Delta < - 1$}}, Phys. Lett. A \textbf{168} (1992),
  256.

\bibitem{JiMi96}
M.~Jimbo and T.~Miwa, \emph{Quantum {KZ} equation with $|q| = 1$ and
  correlation functions of the {XXZ} model in the gapless regime}, J. Phys. A
  \textbf{29} (1996), 2923.

\bibitem{KMT99b}
N.~Kitanine, J.~M. Maillet, and V.~Terras, \emph{Correlation functions of the
  {XXZ} {H}eisenberg spin-$\frac{1}{2}$ chain in a magnetic field}, Nucl. Phys.
  B \textbf{567} (2000), 554.

\bibitem{KuRe81}
P.~P. Kulish and {N.\ Yu.\ Reshetikhin}, \emph{Generalized {H}eisenberg
  ferromagnet and the {G}ross-{N}eveu model}, Zh. Eksp. Teor. Fiz. \textbf{80}
  (1981), 214.

\bibitem{SSNT03}
K.~Sakai, M.~Shiroishi, Y.~Nishiyama, and M.~Takahashi, \emph{Third-neighbor
  correlators of a one-dimensional spin-1/2 {H}eisenberg antiferromagnet},
  Phys. Rev. E \textbf{67} (2003), 065101.

\bibitem{SaSh05}
J.~Sato and M.~Shiroishi, \emph{Fifth-neighbour spin-spin correlator for the
  anti-ferromagnetic {H}eisenberg chain}, J. Phys. A \textbf{38} (2005), L405.

\bibitem{SST05}
J.~Sato, M.~Shiroishi, and M.~Takahashi, \emph{Correlation functions of the
  spin-1/2 anti-ferromagnetic {H}eisenberg chain: exact calculation via the
  generating function}, Nucl. Phys. B \textbf{729} (2005), 441.

\bibitem{Takahashi77}
M.~Takahashi, \emph{Half-filled {Hubbard} model at low temperature}, J. Phys. C
  \textbf{10} (1977), 1289.

\bibitem{TsSh05}
Z.~Tsuboi and M.~Shiroishi, \emph{High temperature expansion of the emptiness
  formation probability for the isotropic {H}eisenberg chain}, J. Phys. A
  \textbf{38} (2005), L363.

\end{thebibliography}

\providecommand{\bysame}{\leavevmode\hbox to3em{\hrulefill}\thinspace}
\providecommand{\MR}{\relax\ifhmode\unskip\space\fi MR }
\providecommand{\MRhref}[2]{%
  \href{http://www.ams.org/mathscinet-getitem?mr=#1}{#2}
}
\providecommand{\href}[2]{#2}

\end{document}